 \documentclass[twocolumn,pra,showpacs,floatfix,amsmath,amssymb,aps]{revtex4-1}
\usepackage{graphicx} 
\usepackage{subfigure}
\usepackage{dcolumn}
\usepackage{bm}
\usepackage{tabularx}
\RequirePackage{rotating}
\usepackage[usenames]{color}

\graphicspath{.}

\newcommand{\la}{\left\langle}
\newcommand{\ra}{\right\rangle}

\newcommand{\ie}{\emph{i.e.}\ }
\newcommand{\etal}{\emph{et al.}\ }
\newcommand{\EA}{^e\!\!A}

\newcommand{\beq}{\begin{equation}}
\newcommand{\eeq}{\end{equation}}
\newcommand{\beqa}{\begin{eqnarray}}
\newcommand{\eeqa}{\end{eqnarray}}
\newcommand{\beqn}{\begin{equation*}}
\newcommand{\eeqn}{\end{equation*}}
\newcommand{\beqan}{\begin{eqnarray*}}
\newcommand{\eeqan}{\end{eqnarray*}}

\newcommand{\ar}{\begin{array}}
\newcommand{\ear}{\end{array}}
\newcommand{\bc}{\begin{color}}
\newcommand{\ec}{\end{color}}
\newcommand{\bit}{\begin{itemize}}
\newcommand{\eit}{\end{itemize}}

\begin{document}

\title{Theoretical study of the isotope effects on the detachment thresholds of Si$^-$}

\author{T. Carette}\email{tcarette@ulb.ac.be}
\author{M. R. Godefroid}\email{mrgodef@ulb.ac.be}
\address{Chimie quantique et photophysique, CP160/09, Universit\'e Libre de Bruxelles, B 1050 Brussels, Belgium}

\date{\today}

\begin{abstract}

The isotope effects in Si$^-$ bound levels are studied using the multi-configuration Hartree-Fock \emph{ab initio} approach. Large scale calculations are carried out for the $3p^3\ ^4S^o,\, ^2D^o$ and $^2P^o$ multiplets of Si$^-$ and the $3p^2\ ^3P$ multiplet of Si.
We predict an anomalous isotope shift on the electron affinity, dominated by the specific mass shift, with a value of $IS(\EA)= -0.66(6)$~m$^{-1}$ for the ($30-28$) isotope pair.
We also report hyperfine structure parameters for the studied multiplets. Finally, we provide the values of level electric field gradients at the nucleus that could be of interest in a study of the metastable silicon isotopes.  Relativistic corrections are estimated using non-relativistic orbitals in the Breit-Pauli and fully relativistic frameworks.

\end{abstract}
\pacs{32.10.Hq, 31.15.aj, 31.30.Gs}



\maketitle

%
\section{Introduction}
%

In the last decades, the interest in the isotope effects in negative ions has grown as the experimental techniques evolved \cite{And:04a,Peg:04a}.
In particular, the isotope shift on the electron affinity, \ie the shift of a negative ion binding energy from one isotope to another, became gradually accessible experimentally \cite{Beretal:95a,Bloetal:01a,Caretal:10a} and theoretically \cite{Caretal:10a,GodFro:99a,CarGod:11b,Carette2013}. The study of isotope shifts on atomic transitions is a rather old subject and previous advances in our understanding of atomic structure are tightly linked to advances in experimental techniques permitting the measurement of isotope effects \cite{BauCha:76a}. The laser photodetachment techniques attained such a level of accuracy that new possibilities for understanding negative ions and isotope effects are now open.

Silicon is the third-period atom of the carbon-group; its lowest configuration is [Ne]$3s^23p^2$. The silicon negative ion binds the three multiplets arising from the $3p^3$ configuration, \ie the ground state $^4S^o_{3/2}$, and the excited $^2D^o_{3/2,5/2}$ and $^2P^o_{1/2,3/2}$ states. \mbox{Scheer \etal \cite{Schetal:98a}} have measured the binding energy of the $^2D^o_{3/2}$ and $^2D^o_{5/2}$ states, 0.527234(25)~eV and 0.525489(20)~eV respectively, but were not able to detect the weakly bound $^2P^o$ which best binding energy measurement to date is due to Kasdan~\etal of 29(5)~meV \cite{Kasetal:75a}. Blondel~\etal \cite{Bloetal:05a} and  Chaibi \etal \cite{Chaetal:10a} later measured the $^{28}$Si electron affinity ($\EA$), \ie the $^4S_{3/2}^o -\ ^3P_0$ threshold, using the Laser Photodetachment Microscopy technique and obtained  $\EA(^{28}$Si$^-)=1~120~724.4(6)$~m$^{-1} = 1.389~5210(7)$~eV.

There are two stable isotopes of silicon with zero spin: the $^{28}$Si ($92.23\%$) and the $^{30}$Si ($3.10\%$). The third stable silicon isotope is the $^{29}$Si ($4.67\%$) and has a spin $I=1/2$.
Lee and Fairbank studied experimentally the $3s^23p^2~^3P_{2} \rightarrow 3s3p^3~^3D_3^o$ transition isotope shifts and, in the case of $^{29}$Si, its hyperfine structure \cite{LeeFai:10a}. It was motivated by the possibility of using the metastable $^{31}$Si ($I=3/2$), decaying by $\beta$ radiation into $^{31}$P, for quantum computing applications \cite{Kan:98a}. Incidentally, it is also the first determination of the hyperfine constant of a state belonging to the ground multiplet of a silicon isotope. Wendt~\etal \cite{Wendt2013} also conducted a two-photon, doppler free study of isotope effects on the $3s^23p^2~^3P_{0,1,2} \rightarrow 3s^23p4p~^3P_{0,1,2}$.

In the present work, we use a similar approach as for previous studies of the $IS$ on the $\EA$ in neighboring elements: sulfur \cite{Caretal:10a}, and chlorine \cite{Carette2013}. This method, relying on a systematic reduction of the single and double excitations of a set of reference configurations has been proven to work efficiently for computing isotopes shifts as well as hyperfine structure parameters \cite{Jonetal:96a}, despite the strong emphasis that this approach puts on providing accurate energies. It has also been successfully used for studying the weakly bound $2p^3\ ^2D^o$ excited state of C$^-$ \cite{CarGod:11b}. With respect to C$^-$, the challenge in Si$^-$ is to correctly describe the correlation of the outer electron with the larger $1s^22s^22p^6$ core. This has been proven to be the bottleneck in S$^-$ and Cl$^-$ studies. One problem is to obtain a balanced description of the neutral atom and negative ion. Following previous works \cite{Caretal:10a,CarGod:11a,Carette2013}, we solve this issue by using orbitals specifically optimized for valence correlation to describe core-valence correlation.

In Section~\ref{sec2}, we briefly lay out the theoretical background. The calculations of the isotope shifts and hyperfine parameters, and their reliability, are detailed in Sections \ref{sec3:methSi} and \ref{sec4:methSi}, respectively. We conclude in Section~\ref{sec5}.

\section{Theory}\label{sec2}

\subsection{Mass isotope shift}\label{sec2:MS}

At the non-relativistic level, the energy corrected for the first order mass shift is \cite{Kin:84a,Carette2013}
\beq
\label{eq:MS}
\delta E^{\small M'M} = \left[ \frac{\mu}{M} - \frac{\mu'}{M'} \right] \left( E_\infty - \frac{\hbar^2}{m_e} S_{sms} \right)
\eeq
where $\mu=m_eM/(m_e+M)$ is the electron reduced mass, $m_e$ the electron mass and $M$ is the bare nucleus mass. $E_\infty$ 
is the total binding energy of the atomic system and $S_{sms}$ is the specific mass shift parameter, both  calculated with an infinite nucleus mass,
\beq
S_{sms} = -\la \Psi_\infty \left| \sum_{i<j} \bm{\nabla}_i \cdot \bm{\nabla}_j \right| \Psi_\infty \ra\ .
\eeq
The first term of (\ref{eq:MS}) contains the normal mass shift (NMS) and the second, the specific mass shift (SMS). The atomic masses are taken from Audi \etal \cite{Audetal:03a}.

\subsection{Field isotope shift}\label{sec2:FS}

It was shown for sulfur \cite{Caretal:10a} and chlorine \cite{CarGod:11a} that even if the field shift (FS) on the electron affinity due to the effect of the finite nucleus volume on the energy levels is below the current experimental resolution, it may constitute a non negligible correction to the total isotope shift on the electron affinity of $p-$block atoms.
This shift can be estimated using
\begin{eqnarray}
\delta E^{\small M'M}_\text{FS} = \frac{ha_0^{3}}{4Z} f(Z)^{\small M'M} \left[ \langle r^2 \rangle_{M} - \langle r^2 \rangle_{M'} \right] 4 \pi \Delta \rho({\bf 0})
\end{eqnarray}
where $f(Z)$ is a scaling factor correcting for the relativistic effects, $\langle r^2 \rangle$ is the isotope-dependent rms radius and $\Delta \rho$ is the change in the spin-less total electron density \cite{Boretal:10a} at the origin
\begin{equation}
\Delta \rho({\bf 0}) = \rho_\text{Si} ({\bf 0}) - \rho_\text{Si$^-$} ({\bf 0})\, .
\end{equation}

The mean square radii of the nucleus charge densities of the different stable isotopes of Si ($A=28,29,30$) are reviewed in Refs. \cite{Ang:99a,FriHei:04a}, offering a large choice of nuclear shape parameters for silicon. We therefore choose to estimate the field shift from the averaged values of Angeli \cite{Ang:04a}, $\la r^2 \ra^{1/2}= 3.1223(24), 3.1168(50), 3.1332(40)$~fm respectively for $A=28,29,30$. The value for $f(Z)/c =1.1099$~m$^{-1}/$fm$^2$ is taken from Aufmuth \etal \cite{Aufetal:87a}. 

\subsection{Hyperfine interaction}\label{sec2:HFS}

The hyperfine structure of a $J$-level is caused by the interaction of the angular momentum of the electron cloud ($\textbf{J}$) and of the nucleus ($\textbf{I}$), forming the total atomic angular momentum $\textbf{F}= \textbf{I} + \textbf{J}$.
The theory underlying the computation of hyperfine structures can be found in Refs \cite{LinRos:74a,Hib:75a,Jonetal:93a,Jonetal:96c}. 
The diagonal hyperfine interaction energy correction is usually expressed in terms of the hyperfine magnetic dipole ($A_J$) and electric quadrupole ($B_J$) constants expressed in MHz. It is possible to further decompose the non-relativistic hyperfine interaction in terms of the $J$-independent orbital~($a_{l}$), spin-dipole~($a_{sd}$), contact~($a_{c}$) and electric quadrupole ($b_{q}$) electronic hyperfine parameters defined in Refs. \citep{LinRos:74a,Hib:75a}.

\subsection{The MCHF expansion}\label{sec2:MCHF}

The multiconfiguration Hartree-Fock (MCHF) approach consists in variationally solving the time-independent Schr\"odinger equation in the space defined by the ansatz \cite{Froetal:97a}
\begin{equation}
\label{eq:SOC}
\Psi(\gamma LS M_L M_S\pi) = \sum_{i} c_i \Phi(\gamma_i LSM_L M_S\pi),
\end{equation} 
where $\Phi(\gamma_i LSM_L M_S\pi)$ are configuration state functions (CSF) built on orthonormal one-electron radial functions.
In practice, we mostly use a Multi-Reference Interacting scheme (MR-I) \cite{Caretal:10a}. It consists in selecting in the expansion~(\ref{eq:SOC}), the CSFs that interact to first order with a Multi-Reference CSF set. This MR-I space is defined in a given one-electron orbital basis set $\lceil n_{max} l_{max} \rceil$ containing in total ($n_{max} - l$) orbitals of angular momentum quantum number $l \leq l_{max}$.
In order to include higher order correlation effects, the linear problem in larger CSF spaces is solved by optimizing the $c_i$ only%
. We refer to this model as configuration interaction (CI) calculations.

Because the specific mass shift on the electron affinity is mainly sensitive to valence correlation, and the hyperfine interaction constants are sensitive to core correlation, we settle for different approaches in Sections \ref{sec3:methSi} and \ref{sec4:methSi}.

All non-relativistic calculations, including the ones of the isotope shift parameters, are performed using the {\sc atsp2k} package \cite{froetal:07a}.

\subsection{Relativistic corrections}\label{sec:rel}

In order to estimate relativistic corrections, we compare non-relativistic calculations to the corresponding relativistic calculations that have similar variational contents. For doing so, the relativistic ansatz
\beq
\Psi(\Gamma JM\pi) = \sum_i c_i \Phi(\gamma_iJM\pi)
\eeq
 are constructed on orbitals optimized at the non-relativistic level. This has been proven to work for hyperfine structures of second period atoms \cite{Jonetal:10a,CarGod:11b,Carette2013e}. For third period atoms, no attempt has been made to assess the reliability of this scheme so far. It has been used for estimating relativistic effects on the hyperfine structures of the ground states of S, S$^-$ and Cl \cite{CarGod:11a}, but this study is not conclusive on the accuracy of the computed corrections.
We compare the Breit-Pauli Configuration Interaction method (BPCI) \cite{Froetal:97a}, and the Relativistic configuration interaction method using the Pauli approximation (RCI-P) \cite{Jonetal:07a}. Such a comparison has recently been performed for excited states of fluorine \cite{Carette2013e}, showing a good consistency between the two approaches.

For differential effects like the electron affinity and its isotope shift, one has to strike a balance in the non-relativistic approach as well as in the relativistic one. Except in the case of carbon and its negative ion \cite{CarGod:11b}, no attempt to achieve this within our framework has been successful. However, as previously emphasized for systems in which it is unrealistic to consider series of calculations converging toward an exact solution \cite{Caretal:10a}, it is necessary to define some guideline to assess the balance of the calculations performed on the neutral and the negative ion. We use the electron affinity itself as the natural guideline. The specific mass shift being much more sensitive to correlation effects than the energy, it is necessary to subtract relativistic corrections on the reference electron affinity, even if relativistic corrections on the isotope shift are not considered. When the nuclear spin is zero, non-relativistic results for the electron affinity can be compared to reference non-relativistic binding energies ($\EA_{ref}^{NR}(LS)$), which are obtained by averaging the fine structure experimental thresholds on the electronic $J$ angular momenta and subtracting a theoretical estimation of the scalar relativistic effects ($\Delta E^{NF}_{th}$) \cite{CarGod:11a}
\beq
\EA_{ref}^{NR}\ =\ \EA_{exp}^{AV} - \Delta E^{NF}_{th}.
\eeq 

The BPCI and  RCI-P calculations are performed using the {\sc atsp2k} package \cite{froetal:07a} and the {\sc grasp2k} package \cite{Jonetal:07a} respectively.

\section{Detachment thresholds and their isotope shifts}\label{sec3:methSi}

We perform HF frozen-core valence ($n=3$) MCHF calculations on the Si $^{3}P$ and Si$^-$ $^4S^o,\, ^2D^o,\,  ^2P^o$ states. Fully variational valence MCHF calculations are also carried out for the Si $^{3}P$ and Si$^-$ $^4S^o$.

We use a similar MR-I approach as in previous works \cite{Caretal:10a,Carette2013}. For Si$^-$, the MR is
\beq
\textrm{MR} = \textrm{[Ne]} \{3s,3p\}^3\{3,4\}^2\quad ^4S^o\, ,\, ^2D^o\, ,\, ^2P^o.
\eeq
where the Ne-like core is kept closed, two electrons are allowed to be excited in correlation orbitals $\{3d,4s,4p,4d,4f\}$ and the three remaining valence electrons are distributed among the spectroscopic $\{3s,3p\}$ orbitals.
More flexibility has to be given to the negative ion model, since it is a system containing one more electron than the neutral atom. An all-electron series of calculations converging toward the exact wave-functions for the anion and corresponding neutral, as performed for instance in the case of Carbon \cite{CarGod:11b}, is intractable in the present case. In general, the best one can hope for when performing \emph{ab initio} calculations is to narrow down an interval in which the targeted property most likely lies by tailoring computational models to the task at hand. It is therefore necessary to use guidelines for assessing the robustness of the error bars.
Hence, for the neutral silicon atom, we choose two multi-reference expansions defined as follows
\beqa
\label{eq:MRSi}
\textrm{MR1}&=& \textrm{[Ne]} \{3s,3p\}^2\{3\}^2\ ^3P \\\label{eq:MR2Si}
\textrm{MR2}&=& \textrm{[Ne]} \{3s,3p\}^2\{3\}^1\{3,4\}^1\ ^3P \, .
\eeqa
We further generate the full \mbox{MR-CV-I$\lceil 10k \rceil$} sets using the above multi-references and allowing at most one hole in the $n=2$ shell.
As advocated in Ref. \cite{Carette2013}, we use the frozen-core  $\lceil 10k \rceil$ orbital basis sets in open-core CI calculations. 
The MR-CV-I expansion of the $^2D^o$ is however too large to be tractable.

Our results for the total energy and $S_{sms}$ parameter of the investigated states are reported in Tables~\ref{tab8:Siparam} and \ref{tab8:Si-param}.

\begin{table}
\caption{\label{tab8:Siparam}Total energies ($E$) and $S_{sms}$ parameters of the $3p^2\ ^3P$ state of silicon calculated by closed-core MCHF and open-core CI calculations. We use two different multi-references, MR1 and MR2, see Eq.~(\ref{eq:MRSi}) and (\ref{eq:MR2Si}). The energies are given in hartrees (E$_\text{h}$) and $S_{sms}$ in units of $a_0^{-2}$.}
\begin{ruledtabular}
\begin{tabular}{cD{.}{.}{6}D{.}{.}{5}cD{.}{.}{6}D{.}{.}{5}}
&\multicolumn{2}{c}{MR1}&&\multicolumn{2}{c}{MR2}  \\
\cline{2-3}\cline{5-6}
$nl$&\multicolumn{1}{c}{$E$}&\multicolumn{1}{c}{$S_{sms}$}&&\multicolumn{1}{c}{$E$}&\multicolumn{1}{c}{$S_{sms}$}\\
&\multicolumn{5}{c}{Frozen-core MR-I, MCHF}\\
$4f$ &-288.936115 & -44.70051 && -288.936207 & -44.70080\\
$5g$ &-288.938949 & -44.69585 && -288.939102 & -44.69631\\
$6h$ &-288.939673 & -44.69836 && -288.939841 & -44.69886\\
$7i$ &-288.939967 & -44.70029 && -288.940139 & -44.70087\\
$8k$ &-288.940104 & -44.70062 && -288.940278 & -44.70121\\
$9k$ &-288.940156 & -44.70064 && -288.940331 & -44.70124\\
$10k$&-288.940176 & -44.70067 && -288.940351 & -44.70126\\
&\multicolumn{5}{c}{Relaxed-core MR-I, MCHF}\\
$10k$&-288.940236 & -44.70410 && -288.940410 & -44.70461\\
&\multicolumn{5}{c}{HF core MR-CV-I, CI}\\
$10k$&-288.974565 & -44.47343 && -288.975221 & -44.47000\\
\end{tabular}
\end{ruledtabular}
\end{table}

\begin{table*}
\caption{\label{tab8:Si-param}Total energies ($E$) and $S_{sms}$ parameters of all bound states of Si$^-$ $3p^3$ calculated by closed-core MCHF and open-core CI calculations. The energies are given in hartrees (E$_\text{h}$) and $S_{sms}$ in units of $a_0^{-2}$.}
\begin{ruledtabular}
\begin{tabular}{cD{.}{.}{6}D{.}{.}{5}cD{.}{.}{6}D{.}{.}{5}cD{.}{.}{6}D{.}{.}{5}}
&\multicolumn{2}{c}{$^4S^o$}&&\multicolumn{2}{c}{$^2D^o$}  &&\multicolumn{2}{c}{$^2P^o$}  \\
\cline{2-3}\cline{5-6}\cline{8-9}\\
$nl$&\multicolumn{1}{c}{$E$}&\multicolumn{1}{c}{$S_{sms}$}&&\multicolumn{1}{c}{$E$}&\multicolumn{1}{c}{$S_{sms}$}&&\multicolumn{1}{c}{$E$}&\multicolumn{1}{c}{$S_{sms}$}\\
&\multicolumn{6}{c}{Frozen-core MR-I, MCHF}\\
$4f$  & -288.985203 & -44.77505 && -288.949710 & -44.75223 && -288.928666 & -44.74459\\
$5g$  & -288.990582 & -44.77066 && -288.957542 & -44.75069 && -288.938660 & -44.74419\\
$6h$  & -288.991886 & -44.77157 && -288.959439 & -44.74967 && -288.941004 & -44.74002\\
$7i$  & -288.992360 & -44.77456 && -288.960126 & -44.75275 && -288.941837 & -44.74061\\
$8k$  & -288.992579 & -44.77503 && -288.960372 & -44.75362 && -288.942200 & -44.74286\\
$9k$  & -288.992662 & -44.77525 && -288.960577 & -44.75402 && -288.942338 & -44.74311\\
$10k$ & -288.992698 & -44.77524 && -288.960631 & -44.75418 && -288.942397 & -44.74330\\
&\multicolumn{6}{c}{Relaxed-core MR-I, MCHF}\\
$10k$ & -288.992749 & -44.77487 & \\
&\multicolumn{6}{c}{HF core MR-CV-I, CI}\\
$10k$ & -289.026801 & -44.54629 && & && -288.975331 & -44.52808 \\
\end{tabular}
\end{ruledtabular}
\end{table*}

\begin{table*}
\caption{Theoretical energies, specific mass shifts (SMS), total mass shifts (MS), field shifts (FS), and total isotope shifts ($IS$) on the detachment thresholds of Si$^-$ for the ($30-28$) isotopic pair. For the $^3P-\ ^4S^o$ valence MCHF calculations, we present the results of both HF frozen core ([Ne] HF) and relaxed core approaches.
All values in $\mbox{cm}^{-1}$.
\label{tab8:SiIS}}
\begin{ruledtabular}
\begin{tabular}{llD{.}{.}{9}D{.}{.}{9}D{.}{.}{9}D{.}{.}{9}}
   & \multicolumn{1}{c }{\hspace*{0.cm} $\EA^{NR}$ \hspace*{0.cm} } 
        & \multicolumn{1}{c }{\hspace*{0.cm} SMS \hspace*{0.cm} }
        & \multicolumn{1}{c }{\hspace*{0.cm} MS\footnote{The NMS is taken from experiment: $0.014647$, $0.005558$ and $0.000306(53)$~cm$^{-1}$ for the $^4S^o$, $^2D^o$ and $^2P^o$ thresholds, respectively. }  \hspace*{0.cm} }  
        & \multicolumn{1}{c }{\hspace*{0.cm} FS \hspace*{0.cm} }  
        & \multicolumn{1}{c }{\hspace*{0.cm} $IS$\footnotemark[1] \hspace*{0.cm} } \\
        \hline
 &\multicolumn{5}{c}{$^3P -\ ^4S^o$}\\
valence relaxed & $11~506(19)$ & -0.02023(8) & -0.00558(8) & 0.00010(6) & -0.00548(14)\\
valence [Ne] HF & $11~508(20)$ & -0.02130(9) & -0.00666(9) & 0.00010(6) & -0.00656(15) \\
+ core-valence  & $11~392(72)$ & -0.0214(5)  & -0.0067(5)  & 0.00010(7) & -0.0066(6) \\
\\
Other theory\footnote{Non-relativistic results from Ref. \cite{Olietal:99a}. Their scalar relativistic correction yields $\EA_{ref}^{NR}=11~420$~m$^{-1}$.}  & $11~425$ \\
$\EA_{ref}^{NR}$   & $11~432$ & \\
\hline
 &\multicolumn{5}{c}{$^3P -\ ^2D^o$}\\
valence & $\phantom{1}4~470(19)$ & -0.01526(9) & -0.00971(9) & 0.00007(5) & -0.00963(14) \\
\\
$\EA_{ref}^{NR}$ & $\phantom{1}4~454$ \\ 
\hline
 &\multicolumn{5}{c}{$^3P -\ ^2P^o$}\\
 valence      & $\phantom{11~}468(20)$ & -0.01214(9) & -0.01184(15) & 0.00006(4) & -0.01178(19) \\
\\
$\EA_{ref}^{NR}$ & $\phantom{11~}438(40)$ \\
\end{tabular}
\end{ruledtabular}
\end{table*}

%
%
%

With the experimental fine structure of the neutral atom, we obtain the $J$-averaged electron affinity
\beq
\EA_{exp}^{AV}(^4S^o)=11~356.93~\textrm{cm}^{-1}.
\eeq
For estimating the scalar relativistic shift, we perform Dirac-Fock calculations with {\sc grasp2k} \cite{Jonetal:07a} and compare them to Hartree-Fock results. We obtain $\Delta E^{NF}_{th}(^4S^o)=-75.1 $~cm$^{-1}$ and
\beq
\EA_{ref}^{NR}(^4S^o)=11~432~\textrm{cm}^{-1}.
\eeq
Note that de Oliveira \etal \cite{Olietal:99a} obtain $\Delta E^{NF}_{th}(^4S^o)=-63.47 $~cm$^{-1}$ when including correlation effects.
Scheer \etal \cite{Schetal:98a} have measured the $^2D^o$ fine structure at $14.08(20)$ cm$^{-1}$ so that the corresponding $\EA_{exp}^{AV}$ is
\beq
\EA_{exp}^{AV}(^2D^o)=4~393.7(3)~\textrm{cm}^{-1}.
\eeq
We calculate $\Delta E^{NF}_{th}(^2D^o)=-60.05$~cm$^{-1}$ so that
\beq
\EA_{ref}^{NR}(^2D^o) = 4~454~\textrm{cm}^{-1}.
\eeq
The Si$^-(^2P^o)$ detachment threshold is $234(40)$~cm$^{-1}$ \cite{Kasetal:75a}. Its fine structure is unknown. Neglecting the possible effect of the $^2P^o$ fine structure
 and using the HF-DF value for $\Delta E^{NF}_{th}(^2P^o)=-54.24$~cm$^{-1}$ we deduce
\beq
\EA_{exp}^{AV}(^2P^o)=384(40)~\textrm{cm}^{-1}
\eeq
and 
\beq
\EA_{ref}^{NR}(^2P^o) = 438(40)~\textrm{cm}^{-1}.
\eeq

As explained in Section~\ref{sec:rel}, calculating relativistic corrections on differential effects including inter-electron correlation, is  delicate. By comparing Hartree-Fock results to Dirac-Fock mass shift parameters calculated using the {\sc ris3} program \cite{Naze2013b}, we estimate relativistic corrections smaller than 1\%, and hence neglect them.

The final prediction is the window between the results obtained from the models based on MR1 and MR2.
This interpolation is expected to provide robust error bars since the $\EA$ and $\Delta S_{sms}$ trends in series of calculations are highly correlated \cite{CarGod:11b,Carette2013}.
Table~\ref{tab8:SiIS} presents the $IS$ on the $\EA$ of Si for the ($30-28$) isotope pair.
The uncertainty on the FS is dominated by the uncertainty on the proton distribution $\delta \la r^2\ra$.
We have an overall good agreement of our non-relativistic calculations with the $\EA_{ref}^{NR}$ and the calculation of de~Oliveira~\etal \cite{Olietal:99a}.
One can easily deduce the hyperfine averaged $IS$ for the isotopic pairs involving the $^{29}$Si with these results.
To the contrary of what has been observed in the calculations of isotope shifts on the electron-affinities of sulfur and chlorine, the relaxed-core MCHF calculations disagree with the open-core CI results, the latter being close to the results obtained in frozen-core calculations.
We observe a breakdown of the MR-CV-I approach for the $^2P^o$ detachment threshold. Indeed this model leads to $\EA_{th}^{NR}(^2P^o)=96(72)$~cm$^{-1}$ which, compared to the value of $468(20)$~cm$^{-1}$ obtained using the closed-core MR-I model, reveals unphysical bias in the $^2P^o$ state open-core calculations. It is due to a significant difference of the role of the orbitals and mixing coefficients in the Si$^-(^2P^o)$ and neutral silicon closed-core expansions due to so-called \emph{quasi-symmetries} in the  MCHF energy functional \cite{Car:10a}. This effect was already encountered, but not fully understood, in neutral sulfur calculations \cite{Caretal:10a}.

\section{Hyperfine structure}\label{sec4:methSi}

For computing hyperfine structures, it is not necessary to get a balance between different states, but inner correlation is of crucial importance.
In this context, we opt for a different systematics in the construction of the MCHF ansatz: an all-electron series of MR-I calculations, hereafter referred as ``open-core MCHF''(OC-MCHF) calculations. The multi-reference is itself a closed-core CSF set including all SD excitations of the valence in the $n=3$ layer but omitting the $3s^2 \rightarrow 3d^2$ excitation. In order to avoid too many redundancies in the variational parameters, the core orbitals are kept frozen to their HF shape in all calculations. Single, double and triple (SDT) excitations of the $\{3s^23p^w,3s^13p^w3d^1\}$ ($w=2$ for Si and $w=3$ for Si$^-$) in $\lceil 4f \rceil$ and $\lceil 5g \rceil$ are added  to the $n_{max}=11$ expansions through configuration interaction (CI).

The results for the hyperfine parameters are given in Tables~\ref{tab8:Siparamhfs} to \ref{Si-2Pparamhfs} for Si($^3P$) and Si$^-$ ($^4S^o$, $^2D^o$, and $^2P^o$). For neutral silicon and Si$^-$~$^4S^o$, we also compare these results with calculations performed with the orbitals $\lceil 10k \rceil$ optimized on valence correlation expansions of Section~\ref{sec2:MCHF}, as done in Ref. \cite{CarGod:11a}. In Tables~\ref{tab8:Siparamhfs} and~\ref{tab8:Si-paramhfs}, the calculation ``V'' stands for the valence MCHF calculation, ``$\cup$ CV'' for the core-valence CI calculation (at most one hole in the core) and the CI calculation ``$\cup$ CC'' includes also the double excitations from the core. As analyzed in Ref. \cite{CarGod:11a}, this latter approach yields good results, despite a slower convergence with the number of correlation layers. They are only used as a indicator of the quality of the results.

\begin{table}
\caption{\label{tab8:Siparamhfs}  Hyperfine parameters, in units of $a_0^{-3}$, obtained from the OC-MCHF and CI calculations performed for the silicon $3p^2~^3P$ term.}
\begin{ruledtabular}
\begin{tabular}{cD{.}{.}{5}D{.}{.}{5}D{.}{.}{5}D{.}{.}{5}}
&\multicolumn{4}{c}{Si $^3P$}\\
$nl$&\multicolumn{1}{c}{$a_{l}$}&\multicolumn{1}{c}{$a_{sd}$}&\multicolumn{1}{c}{$a_c$}&\multicolumn{1}{c}{$b_q$}\\
\cline{2-5}\\
$4f $ & 2.42451 & 0.49677 & 1.84461 & 0.99652\\
$5g $ & 2.42181 & 0.49928 & 1.08012 & 0.97866\\
$6h $ & 2.42420 & 0.50223 & 0.41961 & 0.99904\\
$7i $ & 2.36871 & 0.49710 & 0.57705 & 0.94872\\
$8k $ & 2.36633 & 0.49087 & 0.77783 & 0.95023\\
$9k $ & 2.37003 & 0.49268 & 0.62625 & 0.96203\\
$10k$ & 2.36975 & 0.49204 & 0.69278 & 0.95866\\
$11k$ & 2.36960 & 0.49224 & 0.66352 & 0.96032\\
&\multicolumn{4}{c}{$\cup$ MR-SDT$\lceil n'l' \rceil$, CI} \\
$4f$ & 2.37499 & 0.49344 & 0.61067 & 0.96516\\
$5g$ & 2.37663 & 0.49385 & 0.57875 & 0.96640\\
\hline
&\multicolumn{4}{c}{MR-I$\lceil 10k \rceil$} \\
V& 2.00503 & 0.41095 & 0.83596 & 0.78085 \\
$\cup$ CV& 2.41588 & 0.49439 & 1.83006 & 0.98781 \\
$\cup$ CC& 2.36143 & 0.48519 & 0.62423 & 0.96704
\end{tabular}
\end{ruledtabular}
\end{table}
\begin{table}
\caption{\label{tab8:Si-paramhfs}Hyperfine parameters, in units of $a_0^{-3}$, obtained from the OC-MCHF and CI calculations performed for the Si$^-$ $3p^3~^4S^o$ and $^2D^o$ terms.}
\begin{ruledtabular}
\begin{tabular}{cD{.}{.}{5}ccD{.}{.}{5}D{.}{.}{5}D{.}{.}{5}D{.}{.}{5}}
&\multicolumn{1}{c}{Si$^-$ $^4S^o$}&&&\multicolumn{4}{c}{Si$^-$ $^2D^o$}  \\
$nl$&\multicolumn{1}{c}{$a_c$}&&&\multicolumn{1}{c}{$a_{l}$}&\multicolumn{1}{c}{$a_{sd}$}&\multicolumn{1}{c}{$a_c$}&\multicolumn{1}{c}{$b_q$}\\
\cline{2-3}\cline{5-8}\\
$4f $ & 1.57325 &&& 3.48007 & 0.71108 & 0.53265 & 0.00085\\
$5g $ & 0.89446 &&& 3.47959 & 0.71891 & 0.32564 & 0.00492\\
$6h $ &-0.47211 &&& 3.38189 & 0.70974 &-0.10365 & 0.02545\\
$7i $ &-0.18712 &&& 3.39307 & 0.71769 &-0.01380 & 0.02526\\
$8k $ & 0.00010 &&& 3.39448 & 0.71475 & 0.03436 & 0.02491\\
$9k $ &-0.27006 &&& 3.40187 & 0.71222 &-0.07060 & 0.02242 \\
$10k$ &-0.18106 &&& 3.39954 & 0.71275 &-0.04981 & 0.02365\\
$11k$ &-0.19180 &&& 3.39838 & 0.71207 &-0.05451 & 0.02341\\
&\multicolumn{7}{c}{$\cup$ MR-SDT$\lceil n'l' \rceil$, CI} \\
$4f$ &-0.32457 &&& 3.40735 & 0.71481 &-0.09401 & 0.02533\\
$5g$ &-0.35984 &&& 3.40164 & 0.71508 &-0.10138 & 0.02662\\
\hline
&\multicolumn{7}{c}{MR-I$\lceil 10k \rceil$} \\
V  & -0.00364 \\
$\cup$ CV&  1.05388 \\
$\cup$ CC& -0.24534
\end{tabular}
\end{ruledtabular}
\end{table}
\begin{table}
\caption{\label{Si-2Pparamhfs}Hyperfine parameters, in units of $a_0^{-3}$, obtained from the OC-MCHF and CI calculations performed for the Si$^-$ $3p^3~^2P^o$ term.}
\begin{ruledtabular}
\begin{tabular}{cD{.}{.}{5}D{.}{.}{5}D{.}{.}{5}D{.}{.}{5}}
&\multicolumn{4}{c}{Si$^-$ $^2P^o$} \\
$nl$&\multicolumn{1}{c}{$a_{l}$}&\multicolumn{1}{c}{$a_{sd}$}&\multicolumn{1}{c}{$a_c$}&\multicolumn{1}{c}{$b_q$}\\
\cline{2-5}\\
$4f $ & 1.71550 &-0.35163 & 0.63690 & 0.01448\\
$5g $ & 1.68124 &-0.35213 & 0.28606 &-0.01104\\
$6h $ & 1.68176 &-0.35420 & 0.01997 &-0.01138\\
$7i $ & 1.68515 &-0.35677 & 0.06590 &-0.01515\\
$8k $ & 1.68653 &-0.35637 & 0.10389 &-0.01458\\
$9k $ & 1.68501 &-0.35361 &-0.01197 &-0.01099\\
$10k$ & 1.68502 &-0.35423 &-0.00046 &-0.01227\\
$11k$ & 1.68393 &-0.35370 &-0.00183 &-0.01217\\
&\multicolumn{4}{c}{$\cup$ MR-SDT$\lceil n'l' \rceil$, CI} \\
$4f$ & 1.68787 &-0.35510 &-0.04520 &-0.01535\\
$5g$ & 1.68448 &-0.35850 &-0.04697 &-0.03724
\end{tabular}
\end{ruledtabular}
\end{table}
\begin{table}
\caption{\label{tab:orbVsAc}Mean values and occupation numbers ($q$) of the OC-MCHF$\lceil 11k \rceil$ $s$-orbitals and evolution of the $a_c$ hyperfine parameter along the sequence of OC-MCHF$\lceil nl \rceil$ calculations.}
\begin{ruledtabular}
\begin{tabular}{D{s}{s}{0}D{.}{.}{1}D{.}{.}{3}D{.}{.}{8}D{.}{.}{2}}
  nl  &    \multicolumn{1}{c}{$\langle\delta(r)\rangle$}  &\multicolumn{1}{c}{~~~$\langle r\rangle$}   &      \multicolumn{1}{c}{$q$}    &  a_c\\
\hline
  1s  &     820.8  & 0.111 &  1.99959667  &\\
  2s  &      58.1  & 0.563 &  1.99676579  &\\
  3s  &       3.5  & 2.299 &  1.93575762  &\\
  4s  &     194.8  & 0.845 &  0.00349874  &  1.57\\
  5s  &    2002.7  & 0.400 &  0.00056144  &  0.89\\
  6s  &     130.8  & 2.929 &  0.00318797  & -0.47\\
  7s  &    1646.7  & 0.945 &  0.00009859  & -0.19\\
  8s  &    5183.8  & 0.433 &  0.00002211  &  0.00\\
  9s  &     189.9  & 3.073 &  0.00008889  & -0.27\\
 10s  &    4976.3  & 0.747 &  0.00000190  & -0.18\\
 11s  &    6542.0  & 0.480 &  0.00000029  & -0.19
\end{tabular}
\end{ruledtabular}
\end{table}

\begin{table*}
\caption{\label{tab8:Ahfsrel}Relativistic corrections on $A\frac{I}{\mu_I}$ (MHz per units of $\mu_N$) of each considered state evaluated by comparing SD-MCHF calculations to corresponding RCI-P (RCI) and BPCI (BPCI) results.}
\begin{ruledtabular}
\begin{tabular}{cD{.}{.}{2}D{.}{.}{2}D{.}{.}{2}D{.}{.}{2}cD{.}{.}{2}D{.}{.}{2}cD{.}{.}{2}D{.}{.}{2}D{.}{.}{2}D{.}{.}{2}cD{.}{.}{2}D{.}{.}{2}D{.}{.}{2}D{.}{.}{2}}
&\multicolumn{4}{c}{Si $^3P$}&&\multicolumn{2}{c}{Si$^-$ $^4S^o$} &&\multicolumn{4}{c}{Si$^-$ $^2D^o$} &&\multicolumn{4}{c}{Si$^-$ $^2P^o$} \\
& \multicolumn{2}{c}{$J=1$}& \multicolumn{2}{c}{$J=2$} && \multicolumn{2}{c}{$J=3/2$}&& \multicolumn{2}{c}{$J=3/2$} & \multicolumn{2}{c}{$J=5/2$} && \multicolumn{2}{c}{$J=1/2$} & \multicolumn{2}{c}{$J=3/2$} \\
$nl$& \multicolumn{1}{c}{RCI} & \multicolumn{1}{c}{BPCI}& \multicolumn{1}{c}{RCI} & \multicolumn{1}{c}{BPCI}&& \multicolumn{1}{c}{RCI} & \multicolumn{1}{c}{BPCI}&& \multicolumn{1}{c}{RCI} & \multicolumn{1}{c}{BPCI}& \multicolumn{1}{c}{RCI} & \multicolumn{1}{c}{BPCI}&& \multicolumn{1}{c}{RCI} & \multicolumn{1}{c}{BPCI}& \multicolumn{1}{c}{RCI} & \multicolumn{1}{c}{BPCI}\\
\cline{2-5}\cline{7-8}\cline{10-13}\cline{15-18}\\
$4f$ & -0.48 & -0.52 & 1.72 & 2.05 && -0.37 & -0.25 && 0.35 & 0.34 & 0.67 & 0.93 && 2.47 & 3.27 & 0.21 & 0.32 \\
$5g$ & -1.40 & -1.62 & 2.23 & 2.53 && -1.04 & -1.19 && 0.50 & 0.64 & 1.36 & 1.73 && 3.20 & 4.07 &-0.10 &-0.04 \\
$6h$ & -2.36 & -2.85 & 1.73 & 1.77 && -2.05 & -2.76 && 0.15 & 0.37 & 1.10 & 1.36 && 6.33 & 7.85 &-0.14 &-0.20 \\
$7i$ & -2.49 & -2.98 & 1.77 & 1.89 && -2.21 & -2.97 && 0.22 & 0.48 & 1.21 & 1.52 && 6.85 & 8.60 &-0.16 &-0.21
\end{tabular}
\end{ruledtabular}
\end{table*}

\begin{table*}
\caption{\label{tab8:Bhfsrel}Relativistic corrections on $B/Q$ (MHz/barn) of each considered state evaluated by comparing SD-MCHF calculations to corresponding RCI-P (RCI) and BPCI (BPCI) results.}
\begin{ruledtabular}
\begin{tabular}{cD{.}{.}{2}D{.}{.}{2}D{.}{.}{2}D{.}{.}{2}cD{.}{.}{2}D{.}{.}{2}cD{.}{.}{2}D{.}{.}{2}D{.}{.}{2}D{.}{.}{2}cD{.}{.}{2}D{.}{.}{2}}
&\multicolumn{4}{c}{Si $^3P$}&&\multicolumn{2}{c}{Si$^-$ $^4S^o$} &&\multicolumn{4}{c}{Si$^-$ $^2D^o$} &&\multicolumn{2}{c}{Si$^-$ $^2P^o$} \\
& \multicolumn{2}{c}{$J=1$}& \multicolumn{2}{c}{$J=2$} && \multicolumn{2}{c}{$J=3/2$}&& \multicolumn{2}{c}{$J=3/2$} & \multicolumn{2}{c}{$J=5/2$} && \multicolumn{2}{c}{$J=3/2$} \\
$nl$& \multicolumn{1}{c}{RCI} & \multicolumn{1}{c}{BPCI}& \multicolumn{1}{c}{RCI} & \multicolumn{1}{c}{BPCI}&& \multicolumn{1}{c}{RCI} & \multicolumn{1}{c}{BPCI}&& \multicolumn{1}{c}{RCI} & \multicolumn{1}{c}{BPCI}& \multicolumn{1}{c}{RCI} & \multicolumn{1}{c}{BPCI}&& \multicolumn{1}{c}{RCI} & \multicolumn{1}{c}{BPCI}\\
\cline{2-5}\cline{7-8}\cline{10-13}\cline{15-16}\\
$4f$ & 0.28 & 0.41 & 0.04 &-1.23 && 0.00 &-0.00 && 8.97 & 9.01 & 0.01 & 0.01 &&-8.86 &-8.89 \\
$5g$ & 0.46 & 0.69 &-1.12 &-2.60 &&-0.00 &-0.01 && 8.61 & 8.63 & 0.01 & 0.01 &&-9.40 &-9.41 \\
$6h$ & 0.28 & 0.53 &-1.32 &-2.86 &&-0.00 &-0.01 && 9.18 & 9.17 & 0.03 &-0.01 &&-9.45 &-9.38 \\
$7i$ & 0.28 & 0.57 &-1.39 &-3.01 &&-0.00 &-0.01 && 9.28 & 9.28 & 0.04 & 0.01 &&-9.55 &-9.49 
\end{tabular}
\end{ruledtabular}
\end{table*}

The $a_c$ Fermi contact contribution represents the contact interaction between the nucleus and the electron spins. 
It is well known that this parameter is highly sensitive to spin-polarisation of the electron cloud at the origin and often shows erratic convergence in a sequence of MCHF calculations~\cite{Jonetal:10a}. This difficulty arises from the fact that the relevant CSFs having unpaired $s$-electrons coupled as $(ns ms)^3S$ have very small mixing coefficients $c_i$ in (\ref{eq:SOC}). 
From Tables~\ref{tab8:Siparamhfs}, \ref{tab8:Si-paramhfs} and \ref{Si-2Pparamhfs} only, it is unclear if convergence has been reached for this parameter. 
 The convergence of $a_c$ is especially important for the ground state of the anion, Si$^-$~$^4S^o$, as it is the only non-zero contribution to the magnetic dipole hyperfine constant. 
The most important CSFs for $a_c$ are single excitations ($ns \rightarrow n's$) of the dominant configurations, in particular, to $s$-orbitals with a large contact term $\langle \delta(r) \rangle$. 
In Table~\ref{tab:orbVsAc}, the evolution of the $a_c$ hyperfine parameter calculated with the sequence of OC-MCHF$\lceil nl \rceil$ correlation models, is put in line with the  $\langle \delta(r) \rangle$, mean radius $\langle r \rangle$, and occupation number $q$ of the $s$-orbitals  of the most complete OC-MCHF$\lceil 11k \rceil$ calculation. The orbital reorganization when extending the orbital active set is weak enough to allow a meaningful correlation. Oscillations occur up to $n=8$ but stabilization appears even if the $\langle \delta(r) \rangle_{ns}$ values of the last correlation layers are quite large. Adding higher excitations through MR-SDT has a major effect (see Table~\ref{tab8:Si-paramhfs}) and triple excitations to other layers than $n = 4,5$ might impact $a_c$ even more. 

The non-relativistic electric field gradient at the nucleus $b_q$ of neutral silicon can be estimated to be  accurate to about 0.5~\%. However, like the $a_c$ parameters of Si$^-$~$^4S^o$, $b_q$ is remarkably small for the  Si$^-$ $^2D^o$ and $^2P^o$ multiplets. This is expected from the fact that, within a non-relativistic framework, $b_q$ vanishes for a $p^3$ open-shell.
Relativistic corrections are evaluated by performing MCHF and the corresponding Breit-Pauli and RCI-P calculations on the set of single and double (SD) excitations of the main configuration, as explained in Section~\ref{sec:rel}. The relativistic corrections are the differences between the so-obtained hyperfine constants. They are presented for quantities that are independent of the nuclear parameters, \ie $A\frac{I}{\mu_I}$ in Table~\ref{tab8:Ahfsrel} and $B/Q$ in Table~\ref{tab8:Bhfsrel}. 
There has been no test of which of the two methods is most reliable in this specific context so that we interpret the difference between their results as uncertainties. Overall, the agreement between BPCI and  RCI-P relativistic corrections is satisfactory since they yield uncertainties that are of the same order of magnitude as the degree of convergence of the non-relativistic hyperfine constants.

The $^{29}$Si isotope has a spin $I=1/2$, with a magnetic moment of $\mu(^{29}$Si$)=-0.55529(3)$~$\mu_B$ \cite{Sto:05a}. The calculated $A_J$ hyperfine constants are presented in Table~\ref{tab8:SiHFS}. As a complement of information, the $B/Q$ nuclear-independent constants are also given as they could be useful for the study of metastable isotopes of silicon with non-zero electric quadrupole moment. We compare our results with the experimental value of Lee and Fairbank \cite{LeeFai:10a} for $A_2(^3P)$ and with the constants calculated in the open-core CI approach. This comparison indicate a high degree of convergence of the non-relativistic calculations, \ie to less than 1~\%, except in the case of the small magnetic dipole constants $A_1(^3P)$ of Si and $A_{3/2}(^4S^o)$ of Si$^-$. In the latter case, this is due to the fact that only the problematic fermi-contact term (see the above discussion) contributes to the hyperfine constant. In the case of the $A_1(^3P)$ constant, this relative lack of convergence is due to large cancelation effects.
For a single open shell configuration $l^w\ LSJ$, the ratio between the orbit and spin-dipole contributions to the $A_J$ magnetic dipole hyperfine constant is purely angular
\beqa
\frac{A_J^{dip}}{A_J^{orb}} &=& (-)^{L+S+J+l+1} \frac{g_s}{2}
\sqrt{\frac
	{90 l(l+1)}
	{(2l+3)(2l+2)(2l-1)}}\nonumber\\	
&&\hspace{1.7cm}\left.	\left\{\ar{ccc} L & S & J \\ L & S & J \\ 2 & 1 & 1 \ear \right\}\right/
	\left\{\ar{ccc} L & S & J \\ 1 & J & L \ear \right\}
\eeqa
where $g_s=2.00232$ is the electron gyromagnetic ratio.
For a $p^2$ or $p^4$ open-shell forming a $^3P$, we have
$A_1^{dip}/A_1^{orb} = -\frac{g_s}{2}$, implying that the two contributions cancel each other. This explains why the C, O, Si and S, $A_1(^3P)$ constants are small. However, for higher $Z$ the deviation from the $LS$ coupling increases and the $A_1(^3P)$ constant becomes relatively large~(see \emph{e.g.} Refs. \cite{Oluetal:70a,Chi:71a}).

The drastic effect of relativity on $B_{3/2}/Q$ values in Si$^-$ is striking. Within the non-relativistic approximation, we have $B_{3/2}(^2D^o)/B_{5/2}(^2D^o) = +7/10$. The violation of this relation is due to the small ($0.035~\%$), symmetric $3p^3\ ^2D^o\ -\ 3p^3\ ^2P^o$ mixing at the single-configuration level of approximation. In this (2 x 2) interaction problem, only the cross-term between the $^2D^o$ and $^2P^o$ CSF gives a non-zero $B_{3/2}$. This means that the relativistic corrections to the electric field gradients of the two states are equal in magnitude and of opposite signs, as approximately observed in Table~\ref{tab8:Bhfsrel}. Because the main configuration has an occupation of about 95~\% in both multiplets, the relativistic cross-terms and electron correlation contributions to $b_q$ are of the same order of magnitude.

To complete our work, we report in Table~\ref{tabx:Off_diag_HFS} the  theoretical off-diagonal hyperfine constants~\cite{Jonetal:93a} that affect the splitting in non-zero external magnetic fields where $J$ is no longer a good quantum number.

\begin{table*}
\caption{\label{tab8:SiHFS}$A_J$ and $B_J/Q$ hyperfine constants (in MHz)
calculated with the hyperfine parameters reported in Tables~\ref{tab8:Siparamhfs}, \ref{tab8:Si-paramhfs} and \ref{Si-2Pparamhfs}.
Relativistic corrections and their errors are estimated from Tables~\ref{tab8:Ahfsrel} and~\ref{tab8:Bhfsrel}.}
\begin{ruledtabular}
\begin{tabular}{cD{.}{.}{4}D{-}{-}{7}D{.}{.}{4}D{.}{.}{4}cD{.}{.}{4}c}
&\multicolumn{4}{c}{Si $(^3P)$} && \multicolumn{2}{c}{Si$^-(^4S^o)$}\\
  &  \multicolumn{1}{c}{$A_1$} & \multicolumn{1}{c}{$A_2$}  &  \multicolumn{1}{c}{$B_1/Q$} & \multicolumn{1}{c}{$B_2/Q$} && \multicolumn{2}{c}{$A_{3/2}$}\\
\cline{2-5}\cline{7-8}\\
{\small OC-MCHF}       & -6.73   & -163.38    & 112.82   & -225.64   && 4.52 \\
{\small $\cup$ MR-SDT} & -5.17   & -162.34    & 113.54   & -227.07   && 8.48 \\
{\small +rel}          & -2.1(3) & -164.37(7) & 114.0(2) & -229.3(8) &&11.4(5) \\
\\
exp. \cite{Wendt2013} & \multicolumn{1}{r}{8(10)\phantom{1}} & -163(2) \\
exp. \cite{LeeFai:10a}&       & -160.1(1.3) \\
\\
&\multicolumn{4}{c}{Si$^-(^2D^o)$}\\
& \multicolumn{1}{c}{$A_{3/2}$} & \multicolumn{1}{c}{$A_{5/2}$}& \multicolumn{1}{c}{$B_{3/2}/Q$} & \multicolumn{1}{c}{$B_{5/2}/Q$}\\
\cline{2-5}\\
{\small OC-MCHF}        & -111.07   & -173.25   & -3.85    & -5.50 \\
{\small $\cup$ MR-SDT}  & -111.50   & -173.09   & -4.38    & -6.25 \\
{\small +rel}           & -111.9(2) & -174.6(2) &  4.90(0) & -6.23(2) \\
\\
&\multicolumn{3}{c}{Si$^-(^2P^o)$}\\
& \multicolumn{1}{c}{$A_{1/2}$} & \multicolumn{1}{c}{$A_{3/2}$} & \multicolumn{1}{c}{$B_{3/2}/Q$}\\
\cline{2-4}\\
{\small OC-MCHF}       &  -488.10 & -93.90    &  2.86 \\
{\small $\cup$ MR-SDT} &  -492.63 & -92.53    &  8.75 \\
{\small +rel}          &  -501.2(1.0) & -92.15(3) & -0.77(3)\\
 \end{tabular}
 \end{ruledtabular}
 \end{table*}
 
 \begin{table*}
\caption{\label{tabx:Off_diag_HFS}Off-diagonal $A_{J,J'}$ and $B_{J,J'}/Q$  hyperfine constants (in MHz).}
\begin{ruledtabular}
\begin{tabular}{cD{.}{.}{4}D{-}{-}{7}D{.}{.}{4}D{.}{.}{4}cD{.}{.}{4}c}
&\multicolumn{4}{c}{Si $(^3P)$} && \\
  &  \multicolumn{1}{c}{$A_{1,0}$} & \multicolumn{1}{c}{$A_{2,1}$}  &  \multicolumn{1}{c}{$B_{2,0}/Q$} & \multicolumn{1}{c}{$B_{2,1}/Q$} && \\
\cline{2-5}\\
{\small OC-MCHF}       & -79.26    & -95.86      & -97.71     & -56.41   &&  \\
{\small $\cup$ MR-SDT} & -81.97    & -97.04      & -98.32     & -56.77   && \\
{\small +rel}          & -84.37(3) & -100.78(27) & -98.88(19) & -57.16(14) && \\
\\

\\
&\multicolumn{2}{c}{Si$^-(^2D^o)$} && \multicolumn{2}{c}{Si$^-(^2P^o)$}\\
& \multicolumn{1}{c}{$A_{5/2,3/2}$} & \multicolumn{1}{c}{$B_{5/2,3/2}/Q$} & & \multicolumn{1}{c}{$A_{3/2,1/2}$} & \multicolumn{1}{c}{$B_{3/2,1/2}/Q$} \\
\cline{2-3} \cline{5-6} \\
{\small OC-MCHF}        & -63.22   & -0.60   && -28.25 & 0.62 \\
{\small $\cup$ MR-SDT}  & -64.03   & -0.68   && -28.91 & 1.89 \\
{\small +rel}           & -65.56(19)  & -2.36(1) && -29.76(16) & 2.94(1) \\
 \end{tabular}
 \end{ruledtabular}
 \end{table*}

\section{Conclusion}\label{sec5}

We report the first values of isotope shifts and hyperfine splittings of all bound states of Si$^-$. We also provide the isotope shifts on the binding energy of those states and the hyperfine structure constants of the $3p^2\ ^3P$ lowest multiplet of Si. For the latter, we obtain a satisfactory agreement with experiment. We also find a good consistency between the calculated photodetachment thresholds and their ``non-relativistic experimental" values ($\EA^{NR}_{ref}$), deduced by subtracting the experimental data and the relativistic corrections.

Most hyperfine constants are determined to about $\sim 1\%$. These results could be useful for analyzing experimental spectra where hyperfine structure might not be resolved, but still be significant at the level of the experimental uncertainty, as is the case of recent laser photodetachment microscopy experiments on P$^-$ \cite{Peletal:11a}.

We present the first systematic comparison of configuration interaction relativistic methods based on non-relativistic orbitals (BPCI and RCI-P) for a third period atom. The overall consistency between the so-deduced corrections, in particular in the cases where they account for a large fraction of the hyperfine constants, brings a new evidence that they yield useful estimates of relativistic effects. This is particularly interesting in the context of non-relativistic methods, as it is in general necessary to consider the impact of relativity on the results \cite{Carette2013e}.

\section*{Acknowledgements}
\noindent
This work was supported by the \emph{Communaut\'e fran{\c c}aise of Belgium} (\emph{Action de Recherche Concert\'ee}), the Belgian National Fund for Scientific Research (FRFC/IISN Convention), and the IUAP Belgian State Science Policy (Brix network P7/12).

\bibliography{/Users/thomas/Documents_Tip_Top/library.bib,/Users/thomas/Documents_Tip_Top/Biblio_Thomas.bib}
\end{document}